\documentclass[prd,onecolumn,nofootinbib,aps,tightenlines,preprintnumbers]{revtex4}

\usepackage{amsmath,amssymb}
\usepackage{amsfonts}
\usepackage{color}
\usepackage{graphicx}

\def\be{\begin{equation}}
\def\ee{\end{equation}}
\def\bea{\begin{eqnarray}}
\def\eea{\end{eqnarray}}

\newcommand{\units}[1]{\ensuremath{\mathrm{#1}}}

\newcommand{\mev}{\units{MeV}}
\newcommand{\gev}{\units{GeV}}
\newcommand{\tev}{\units{TeV}}
\newcommand{\cm}{\units{cm}}

\newcommand{\s}{\units{s}}
\newcommand{\sr}{\units{sr}}
\newcommand{\pb}{\text{pb}}
\newcommand{\yr}{\text{yr}}

\graphicspath{ {../Figures/} }

\begin{document}

\preprint{\hbox{PREPRINT UH511-1297-2018}  }

\title{Indirect Detection of Sub-GeV Dark Matter Coupling to Quarks}
\author{Jason Kumar}
\affiliation{Department of Physics and Astronomy, University of Hawai'i,
  Honolulu, HI 96822, USA}

\begin{abstract}
We consider photon signals arising from the annihilation or decay of low-mass (sub-GeV) dark matter which
couples dominantly to quarks.  In this scenario, the branching fractions to the
various kinematically accessible hadronic final states can largely be determined from chiral
perturbation theory.  Several of these final states yield striking spectral features in the sub-GeV photon
spectrum.  New experiments, such as e-ASTROGAM and AMEGO, are in development to improve sensitivity in this
energy range, and we discuss their potential sensitivity to this class of models.
\end{abstract}
\maketitle

\section{Introduction}

There has been much recent interest in dark matter models in which the candidate
particle has a mass $\lesssim {\cal O}(\gev)$.  A variety of theoretical models have
been developed in which such a candidate can have a weak coupling to Standard Model particles
and can obtain a relic density consistent with
observational limits (see, for example,~\cite{Dolgov:1980uu,Carlson:1992fn,Moroi:1993mb,Hall:2009bx,Hochberg:2014dra,Kuflik:2015isi}).
Moreover, the sensitivity of direct detection experiments  tends to be suppressed at such small masses, allowing these
models to escape stringent experimental constraints.  As such, a variety of experiments
have been proposed to improve sensitivity to this class of models.  In particular, a variety
of astrophysical observatories, including e-ASTROGAM~\cite{DeAngelis:2017gra} and
AMEGO~\cite{Caputo:2017sjw}, are being developed to fill in
the current ``MeV-gap" in experimental sensitivity to photons, and these instruments will be
well-positioned to probe the annihilation or decay of sub-GeV dark matter.

It has recently been pointed out that, if sub-GeV dark matter couples predominantly to quarks,
then the indirect detection signatures are particularly striking~\cite{Boddy:2015efa,Boddy:2015fsa,Boddy:2016fds,Bartels:2017dpb,Cata:2017jar,Dutra:2018gmv}.  This is because there are
few kinematically accessible particles  when the center-of-mass energy of the annihilation or
decay process is $\sqrt{s} \lesssim {\cal O}(\gev)$, and those particles tend to yield fairly striking
photon signatures.  In~\cite{Boddy:2015efa,Boddy:2015fsa}, the case $\sqrt{s} < 2m_{\pi^\pm}$ was
considered.  In this case, the dominant two-body final states are $\gamma \gamma$, $\gamma \pi^0$ and
$\pi^0 \pi^0$; the particular final state is determined by the quantum numbers of the initial state, and
the final photon spectra are particularly simple.  If $\sqrt{s} > 2m_{\pi^\pm}$,
then there are typically multiple final states for any choice of the initial state quantum numbers, and
three-body final states are also important.  But the branching fractions and kinematic distributions of the
final state particles can be estimated using chiral perturbation theory (see, for example,~\cite{Cata:2017jar} and~\cite{CHPTReviews}).
In this work, we will derive the
photon spectra which arise in general for dark matter annihilation or decay to light mesons, assuming
$\sqrt{s} \lesssim 1~\gev$.

Our main assumption will be that primary electroweak interactions are negligible; the primary products of
dark matter annihilation or decay will consist only of light mesons, with photons produced only by meson decay.
For this purpose, we will find that the most important final states are those containing an $\eta$, which decays
to $\gamma \gamma$ with $\sim 40\%$ branching fraction.
We will limit ourselves to final states with at most three mesons.  We will assume that dark
matter couples to light quarks, and will find that the available final states can be classified by the quantum
numbers of the initial state.  We will determine the branching fractions and spectra of all relevant final states using chiral
perturbation theory, and will assess the sensitivity of current and upcoming instruments.

The plan of this paper is as follows.  In section II, we will describe the application of chiral perturbation theory
to sub-GeV dark matter.  In section III, we will describe the photon spectra arising from meson decay.  In section IV,
we will present our results, and we conclude in section V.

\section{Applying Chiral Perturbation Theory to the Interactions of sub-GeV dark matter}

We consider the scenario in which low-mass dark matter annihilates or decays via a $C$- and $P$-conserving contact interaction
with light quarks ($u$, $d$, $s$).  We assume $\sqrt{s} < 1~\gev$, and we assume that interactions which scale
as $\alpha_{EM}$ or $G_F$ are negligible (i.e., the dominant primary interaction is QCD).  As a result of these
assumptions, we will find these selection rules:
\begin{itemize}
\item{The charge conjugation ($C$) and parity ($P$) transformation properties of the initial state and final state must be the
same, since $C$ and $P$ are conserved by QCD.  }
\item{The final state must have strangeness equal to zero, since QCD also conserves strangeness.}
\end{itemize}
Subject to these selection rules, the only mesons which we will be interested in are $\pi^0~(m_{\pi^0} \sim 135~\mev)$,
$\pi^\pm~(m_{\pi^\pm} \sim 140~\mev)$ and $\eta~(m_\eta \sim 548~\mev)$.
Since kaons must appear in pairs to conserve strangeness, they can only be produced if $\sqrt{s}$ is very close to
$1~\gev$.  $\rho^\pm$, $\rho^0$ can also be produced in conjunction with pions, but only if $\sqrt{s} \gtrsim 0.91~\gev$.
Although there are some regions of parameter space where these particles can be relevant, they are kinematically inaccessible in
most of the parameter space we consider, so we will ignore them.

The decay of $\pi^\pm$ generally produces muons and neutrinos, with only a small contribution to the
photon spectrum.  The mesons most relevant to the photon spectrum are $\pi^0$, which decays to $\gamma \gamma$ with
a $\sim 99\%$ branching fraction, and $\eta$, which decays to $\gamma \gamma$ with a $\sim 39\%$ branching fraction.  Note,
$\eta$ also decays to $3\pi^0$ with a $\sim 33\%$ branching fraction, and to $\pi^+ \pi^- \pi^0$ with a $\sim 23\%$ branching
fraction.  As a result, a single annihilation or decay process can
produce a relatively large number of neutral pions.  Although this results in a larger number of secondary photons, their energy
spectrum is much more complicated.  Moreover,  the astrophysical gamma ray backgrounds tend to grow rapidly at lower energy; since
the photons arising from $\pi^0$ decay tend to be much less energetic than those arising from $\eta$ decay, they will compete against a
much larger background.  As such, we will focus only on the photons arising from $\eta$ decay.
If $\sqrt{s} < 1~\gev$, then the only kinematically allowed final states including at least one $\eta$ are
$\pi^0 \eta$, $\pi^0 \pi^0 \eta$ and $\pi^+ \pi^- \eta$.

Since the relevant mesons are all pseudo-Nambu Goldstone bosons (pNGBs) of chiral symmetry breaking, their interactions
with dark matter can be described using chiral perturbation theory, in which the dark matter is treated as a
spurion whose interactions break flavor symmetry.  This spurion is set equal to the dark sector operator
which couples to quark bilinears in the fundamental Lagrangian.
For the case of dark matter decay, the spurion will be the dark matter field, whereas for the case of
dark matter annihilation, the spurion will be the appropriate dark matter bilinear, weighted by an energy scale
associated with the energy scale of dark sector interactions.  In the chiral Lagrangian, this spurion then couples
to the octet of pNGBs, and the form of this interaction is determined by the Lorentz, parity, and flavor transformation
properties of the quark bilinears to which the spurion couples in the fundamental Lagrangian.
In the Standard Model, one already introduces scalar, pseudoscalar, vector, and axial-vector spurions in order to
describe quark masses and electroweak interactions.  As a result, one can use data to determine the coefficients of
the operators coupling these spurions to the meson octet, order by order in the chiral Lagrangian.  We will thus only
consider the cases in which dark matter couples to scalar, pseudoscalar, vector, or axial-vector quark currents, and the
corresponding spurions will be denoted as $s$, $p$, $v_\mu$ and $a_\mu$, respectively.

We will work to lowest order in the chiral Lagrangian; this will be a good approximation for our purposes, especially
if the momenta of the final state particles are small.  The effective Lagrangian is thus given by
\bea
{\cal L} &=& \frac{F^2}{4} Tr \left[(D_\mu U D^\mu U^\dagger + \chi U^\dagger + U \chi^\dagger \right] ,
\eea
where
\bea
U &\equiv& \exp [\imath \sqrt{2} \Phi / F] ,
\nonumber\\
\Phi &\equiv& \left(
                \begin{array}{ccc}
                  \frac{\pi^0}{\sqrt{2}} + \frac{\eta_8}{\sqrt{6}} & \pi^+ & K^+ \\
                  \pi^- & -\frac{\pi^0}{\sqrt{2}} + \frac{\eta_8}{\sqrt{6}} & K^0 \\
                  K^- & \bar K^0 & -\frac{2\eta_8}{\sqrt{6}} \\
                \end{array}
              \right) ,
\nonumber\\
\chi &=& 2B (s+\imath p) ,
\nonumber\\
D_\mu U &=& \partial_\mu -\imath (v_\mu + a_\mu)U +\imath U (v_\mu - a_\mu) .
\eea
The field $\eta_8$ appearing above is a linear combination of the physical
$\eta$ and $\eta'$, with $\eta_8 = \eta \cos \theta_P  + \eta' \sin \theta_P $, and $\theta_P \sim 11.5^\circ$
($\cos \theta_P \sim 0.98$).
Henceforth, for simplicity, we will simply equate $\eta_8$ with the physical $\eta$.
As expected, the constants $F$ and $B$ are determined from data;
$F \sim F_\pi \sim 92~\mev$ is the pion decay constant, and $B = m_\pi^2 /(m_u + m_d) + {\cal O}(m^2)$.  Henceforth,
we will take $m_{\pi^0} \sim m_{\pi^\pm} = m_\pi$, as this approximation will only have a non-negligible effect very near
the threshold for $\pi^\pm$ production.

The chirality-violating real-valued spurions are given by
\bea
s &=& \left(
        \begin{array}{ccc}
          m_u + \alpha_S^u \frac{\bar X X}{\Lambda^2} & 0 & 0 \\
          0 & m_d + \alpha_S^d \frac{\bar X X}{\Lambda^2} & 0 \\
          0 & 0 & m_s + \alpha_S^s \frac{\bar X X}{\Lambda^2} \\
        \end{array}
      \right) ,
\nonumber\\
p &=& \left(
        \begin{array}{ccc}
          \alpha_P^u \frac{\imath \bar X \gamma^5 X}{\Lambda^2} & 0 & 0 \\
          0 & \alpha_P^d \frac{\imath \bar X \gamma^5 X}{\Lambda^2} & 0 \\
          0 & 0 & \alpha_P^s \frac{\imath \bar X \gamma^5 X}{\Lambda^2} \\
        \end{array}
      \right) ,
\nonumber\\
v_\mu &=& \left(
        \begin{array}{ccc}
          \alpha_V^u \frac{\bar X \gamma_\mu X}{\Lambda^2} & 0 & 0 \\
          0 & \alpha_V^d \frac{\bar X \gamma_\mu X}{\Lambda^2} & 0 \\
          0 & 0 & \alpha_V^s \frac{\bar X \gamma_\mu X}{\Lambda^2} \\
        \end{array}
      \right) ,
      \nonumber\\
a_\mu &=& \left(
        \begin{array}{ccc}
          \alpha_A^u \frac{\bar X \gamma_\mu \gamma^5 X}{\Lambda^2} & 0 & 0 \\
          0 & \alpha_A^d \frac{\bar X \gamma_\mu \gamma^5 X}{\Lambda^2} & 0 \\
          0 & 0 & \alpha_A^s \frac{\bar X \gamma_\mu \gamma^5 X}{\Lambda^2} \\
        \end{array}
      \right) ,
\eea
where, for simplicity, we have considered the case of dark matter annihilation.  If dark matter
instead decays, the dark matter bilinear would be replaced by a single dark matter field.  The dimensionless coefficients
are equal to the coefficients of the operator coupling the dark matter bilinear to the appropriate quark current in
the fundamental Lagrangian.

We can now expand this Lagrangian, in order to determine the coefficients of the
operators which couple dark matter to the light mesons.  Keeping only the terms which couple the spurions to
at most three mesons, we find
\bea
{\cal L} &=& F^2 B Tr \left[s -\frac{1}{F^2} \left( s\Phi^2 \right) +\frac{\sqrt{2}}{F}   \left(p \Phi \right) - \frac{\sqrt{2}}{3F^3} \left( p\Phi^3 \right) \right]
+\imath Tr \left[v_\mu  \left( (\partial^\mu \Phi ) \Phi    -  \Phi (\partial^\mu \Phi )  \right)  \right]
\nonumber\\
&\,&
+\sqrt{2} F Tr \left[(\partial^\mu a_\mu ) \left(\Phi - \frac{\Phi^3}{3F^2} \right) \right]
-\frac{\sqrt{2}}{ F} Tr \left[ a_\mu  \Phi \left( \partial^\mu \Phi \right)\Phi\right]  +....
\label{eq:TruncatedLagrangian}
\eea
We immediately see that, at this order in the chiral Lagrangian, there are contact interactions which couple $s$  to two-body $L=0$ states
and couple $v_i$ to two-body $L=1$ states.  But the contact interactions couple  $p$ and $a_0$ to three-body $L=0$ states, and couple
$a_i$  to three-body $L=1$ states. But $p$ and $a_\mu$ can also couple directly to $\pi^0$ and $\eta$, allowing dark matter to annihilate
through a mediator in the $s$-channel.  In the case of dark matter decay, we would instead find mixing between dark matter and either $\pi^0$
or $\eta$, and this mixing can be constrained by data.  But this coupling of dark matter to a single meson will vanish if the coefficients are flavor-universal.

A few other features are apparent from the Lagrangian in eq.~\ref{eq:TruncatedLagrangian}.  The contact interactions between dark matter and the final
states we consider are actually independent of the dark matter coupling to strange quarks.  We can see this by noting that all of the terms in
eq.~\ref{eq:TruncatedLagrangian} involve a single trace; thus, if a trace involves more than one insertion of $\Phi$, then $\alpha_{S,P,A,V}^s$
can only appear along with $\Phi_{3i}$ and $\Phi_{j3}$.
These terms are only relevant if the final state has more than one $\eta$ or kaon, and these states
are not kinematically accessible.  To study contact interactions, we may thus truncate $\Phi$ to the upper-left $2 \times 2$ block, which we denote
as $\tilde \Phi = (\eta/\sqrt{6})\sigma_0 + (\pi^0/\sqrt{2})\sigma_3 +\pi^+ \sigma_+ + \pi^- \sigma_- $.  Note, however, that a dark matter coupling
to strange quarks does permit dark matter to annihilate to a three-body final state via an intermediate $\eta$ in the $s$-channel.

As each spurion is a diagonal $3 \times 3$ matrix, it will be convenient to parameterize the flavor structure of
spurions by expressing them as a linear combination of
$M_1 = diag(1,1,1), M_2 = diag(1,-1,0), M_3 = diag(-1,-1,2) $.  In particular, spurions proportional to $M_1$ and $M_3$ will
have the same contact interactions, but the spurion proportional to $M_3$ will also couple to mesons through an intermediate
$\eta$, while the spurion proportional to $M_1$ will have only contact interactions.  A spurion proportional to $M_2$ will have
contact interactions with mesons, as well as a coupling through an intermediate $\pi^0$.

$\tilde \Phi$ is a linear combination of an isospin singlet ($\eta$) and an isospin triplet ($\pi^\pm, \pi^0$), and the isospin of the final state is determined by the isospin of the spurion associated with the initial state.
In particular, if the spurion describing the initial state is
proportional to $M_1$ or $M_3$, then the final state has $I=0$, whereas
if the spurion is proportional to $M_2$, then the final state has $I=1$, $I_3=0$ ($I$ is the isospin quantum number).  It is straightforward to determine which final states
can be produced by various choices of spurions using the transformation properties of various particles under $C$, $P$ and $I$.  Note that a
$\pi \pi$ pair with $I_3=0$ transforms as $C:(-1)^L$, $P:(-1)^L$, where $L$ is the angular momentum
of the two-body state (for example, see~\cite{Kumar:2013iva}).  Symmetry of the wavefunction under particle interchange
then requires $I = L~mod~2$.
We summarize the connection between various spurions and the relevant possible final states in Table~\ref{tab:AllowedFinalStates}.

\begin{table}[h]
\centering
\begin{tabular}{|c|c|c|c|c|}
  \hline
  spurion & $J^{PC}$  & available states & $\#~\eta$ & $\#~\pi^0$ \\
  \hline
  $s^{M_2}$ & $0^{++}$  & $\pi^0 \eta$  & 1 & 1 \\
  $p^{M_1,M_3}$, $a^{M_1,M_3}_0$  & $0^{-+}$ & $\eta (\pi^+ \pi^-, \pi^0 \pi^0)_{I=0} $  &  1& 2/3 \\
  \hline
\end{tabular}
\caption{For each choice of spurion ($s,p,v_\mu,a_\mu$) and flavor structure ($M_1,M_2,M_3$), we list the
$J^{PC}$ quantum numbers of the initial state,  the possible final states, and the average number of primary
$\eta$ and $\pi^0$ particles produced per annihilation/decay event.  For final states with more than one
pion, we indicate the total isospin of the pion state.  We include only spurions which couple to final states
including an $\eta$.}
\label{tab:AllowedFinalStates}
\end{table}

Note that, at lowest order in the chiral Lagrangian, the $v_i$ spurion couples only to the
$\pi^+ \pi^-$ state, whose decays yield few secondary photons.  (The two-body final state must have $J^{PC} = 1^{--}$,
and must thus be a $\pi \pi $-state with $I=L=1$; such a state has no $\pi^0 \pi^0$ contribution.)
Final states involving $\pi^0$ or
$\eta$ arise only at higher order in the momentum expansion, and will have a small branching fraction.
As such, we will henceforth ignore the case in which dark matter couples to vector quark currents.  Similarly,
there are several spurions which only couple to $\pi \pi$ or $\pi \pi \pi$ final states, and we ignore them as
well.

We are left with only the spurions $s^{M_2}$, $p^{M_1,M_3}$ and $a_0^{M_1,M_3}$.  The spurion
$s^{M_2}$ can only produce the final state $\pi^0 \eta$, as this is the only two-body $0^{++}$ final state
with $I=1$, $I_3=0$.  On the other hand the spurions $p^{M_1,M_3}$ and $a_0^{M_1,M_3}$ can each produce
two three-body final states: $\pi^0 \pi^0 \eta$ and $\pi^+ \pi^- \eta$.  The branching fractions to these
states ($2/3$ and $1/3$, respectively) are determined by the fact that the two outgoing pions must be in
an $I=0$ state.  But in any case, the branching fractions are  not really relevant to our analysis, as we are
only focusing on photons arising from the decay of the $\eta$.

Although we have obtained these results from considerations of symmetry, we can of course
verify, by calculating explicitly in the chiral Lagrangian, that only the spurions
$s^{M_2}$, $p^{M_1,M_3}$ and $a_0^{M_1,M_3}$ can produce two- or three-body final states containing an $\eta$.
But it is possible that one of the final states allowed by symmetry is nevertheless not produced at this order
in the chiral Lagrangian, due to an accidental cancellation.  Indeed, it turns out that, at this order in the
chiral Lagrangian, there is no contact interaction coupling the spurions $a_0^{M_1,M_3}$ to the $\eta \pi \pi$ final
state, as the result of an accidental cancellation.

But unlike the spurions $(p,a_0)^{M_1}$, the spurions $(p,a_0)^{M_3}$ can also couple to an intermediate
$\eta$ in the $s$-channel.  Thus, although the spurion $a_0^{M_3}$ does not produce an $\eta \pi \pi $ final
state through a contact interaction, it does produce this final state through a diagram with an intermediate off-shell
$\eta$.  However, this interaction also vanishes for the spurion  $a_0^{M_1}$.  For the  $a_0^{M_1}$ spurion, although the $\eta \pi \pi $
final state is allowed by symmetry, it is not produced at this order in the chiral Lagrangian.  One can verify that
for all other spurions, every relevant final state allowed by symmetry is produced at this order.

\section{Photon Spectra}

Given a particular choice of Lorentz and flavor structure for the dark matter coupling to quarks, it is
straightforward to determine the resulting secondary photon spectrum.  Since the secondary photons arise
from the process $\eta  \rightarrow \gamma \gamma$, the resulting photon spectrum is simply related to
energy spectrum of the parent $\eta$.

\subsection{$\eta$ Injection Spectra}

If dark matter annihilation or decay, with center-of-mass energy $\sqrt{s}$, produces a $\pi^0 \eta$ final state, then the energy spectrum of
the outgoing $\eta$ is given by
\bea
\frac{dN_\eta}{dE_\eta} &=&  \delta \left(\sqrt{m_\eta^2 + \frac{s}{4} \left(1-\frac{m_\eta^2 + m_\pi^2}{s} \right)^2
- \frac{m_\eta^2 m_\pi^2}{s} } \right)
~ \Theta \left(\sqrt{s} -m_\eta -m_\pi \right) .
\eea

If dark matter annihilation or decay produces a $\pi \pi \eta$ final state, then the energy spectrum of the $\eta$
can be determined by integrating the squared matrix element over the three-body phase space,
which can be expressed in terms of the energies of any two of the three final state particles.  In particular,
we have
\bea
d\Phi_3  &=& \frac{s}{128\pi^3}  dx_1  dx_2  ~ f ,
\nonumber\\
f &=&
\Theta \left( 2-x_1-x_2 - \frac{2}{\sqrt{s}} \sqrt{m_3^2
+\left[\sqrt{\frac{s x_1^2}{4} -m_1^2} - \sqrt{\frac{s x_2^2}{4} -m_2^2} \right]^2} \right) ,
\nonumber\\
&\,& \times \Theta \left(x_1+x_2 + \frac{2}{\sqrt{s}} \sqrt{m_3^2
+\left[\sqrt{\frac{s x_1^2}{4} -m_1^2} + \sqrt{\frac{s x_2^2}{4} -m_2^2} \right]^2} -2 \right) ,
\eea
where $x_i \equiv 2 E_i / \sqrt{s}$, $\sum_{i=1}^3 x_i =2$.
Note, if the final state  is $\pi^0 \pi^0 \eta$, then the phase space integral has an
additional $1/2$ combinatoric factor.
We will take  $i=1$ to denote the $\eta$, and $i=2,3$ to denote the two pions
(either $\pi^+ \pi^-$ or $\pi^0 \pi^0$).  Then we find
\bea
\frac{dN}{dx_1} &=& \frac{\int dx_2~ |{\cal M}|^2 f}{\int dx_1~ dx_2 |{\cal M}|^2 f} .
\eea
Note that the combinatoric factor cancels in this expression.  In general, the energy spectrum of
a meson which is part of a three-body final state depends on the energy dependence of the matrix element.
But one can verify from the chiral Lagrangian that, for the spurions we consider here, the matrix element is
independent of $x_i$.  But one can also see this entirely from considerations of symmetry.  The spurions which
can produce an $\eta \pi \pi$ final state have $J^{PC} = 0^{-+}$ and $I=0$.  Isospin conservation thus requires
the $\pi \pi$ two-body state to have vanishing total isospin, and symmetry of the wavefunction then requires the $\pi \pi$
two-body state to have $J^{PC}=L_\pi^{++}$, where $L_\pi = 0~mod~2$ is the orbital angular momentum of the $\pi \pi$ state.
Angular momentum conservation then requires $L_\eta = L_\pi$, where $L_\eta$ is the orbital angular momentum of the $\eta (\pi \pi)$ state.
As a result, the matrix element can only have a non-trivial dependence on the $x_i$ if there are at least 4 derivatives ($L_\eta = L_\pi =2$),
and no such terms can arise at this order in the chiral Lagrangian.

We thus find
\bea
\frac{dN}{dx_1} &=& \frac{\int dx_2~ f}{\int dx_1~ dx_2  f} .
\eea
The injection spectrum for $\eta$ is a single bump, which vanishes at the kinematic
endpoints
\bea
x_{min} &=& \frac{2m_\eta}{\sqrt{s}} ,
\nonumber\\
x_{max} &=& \sqrt{\frac{4m_\eta^2}{s} + \left(1-\frac{m_\eta^2 + 4m_\pi^2}{s} \right)^2
- \frac{16m_\eta^2 m_\pi^2}{s^2} } .
\eea
$x_{min}$ corresponds to the limit in which the $\eta$ is produced with no boost, and the
two pions are produced back-to-back, while $x_{max}$ corresponds to the limit in which
both pions move with the same boost, in the opposite direction to the $\eta$.

\subsection{Secondary Photon Spectrum}

Given the injection spectrum of the parent $\eta$, the secondary photon spectrum is
given by~\cite{Boddy:2016hbp}
\bea
\frac{dN_\gamma}{dE_\gamma} &=& 2 \int_{\frac{m_\eta}{\sqrt{s}}\left(\frac{2E_\gamma}{m_\eta}
+\frac{m_\eta}{2E\gamma} \right)}^\infty
dx_1 \left[\frac{dN_\eta}{dx_1}
\frac{1}{\sqrt{(s x_1^2/4)-m_\eta^2}} \right] .
\label{eq:PhotonSpectrum}
\eea
A few points about the photon spectrum are immediately apparent~\cite{Boddy:2016hbp}.  This spectrum is log-symmetric about
the energy scale $E_* = m_\eta /2$, and has a global maximum at $E_\gamma = E_*$.  Moreover,
the photon spectrum is non-increasing as $E_\gamma$ either increases or decreases away from $E_*$.

If dark matter decay or annihilation produces a $\pi^0 \eta$, then the $\eta$ is
monoenergetic with an energy determined by $\sqrt{s}$; we will denote the boost of the $\eta$ by
$\gamma_\eta = E_\eta / m_\eta$, with $\beta_\eta = \sqrt{1-(1/\gamma_\eta)^2}$.  From eq.~\ref{eq:PhotonSpectrum},
we see that the photon spectrum is
uniform between the limits $(m_\eta/2) [\gamma_\eta (1-\beta_\eta)]$ and $(m_\eta/2) [\gamma_\eta (1+\beta_\eta)]$,
and vanishes outside these limits.

\begin{figure}[t]
  \centering
  \includegraphics[scale=0.80]{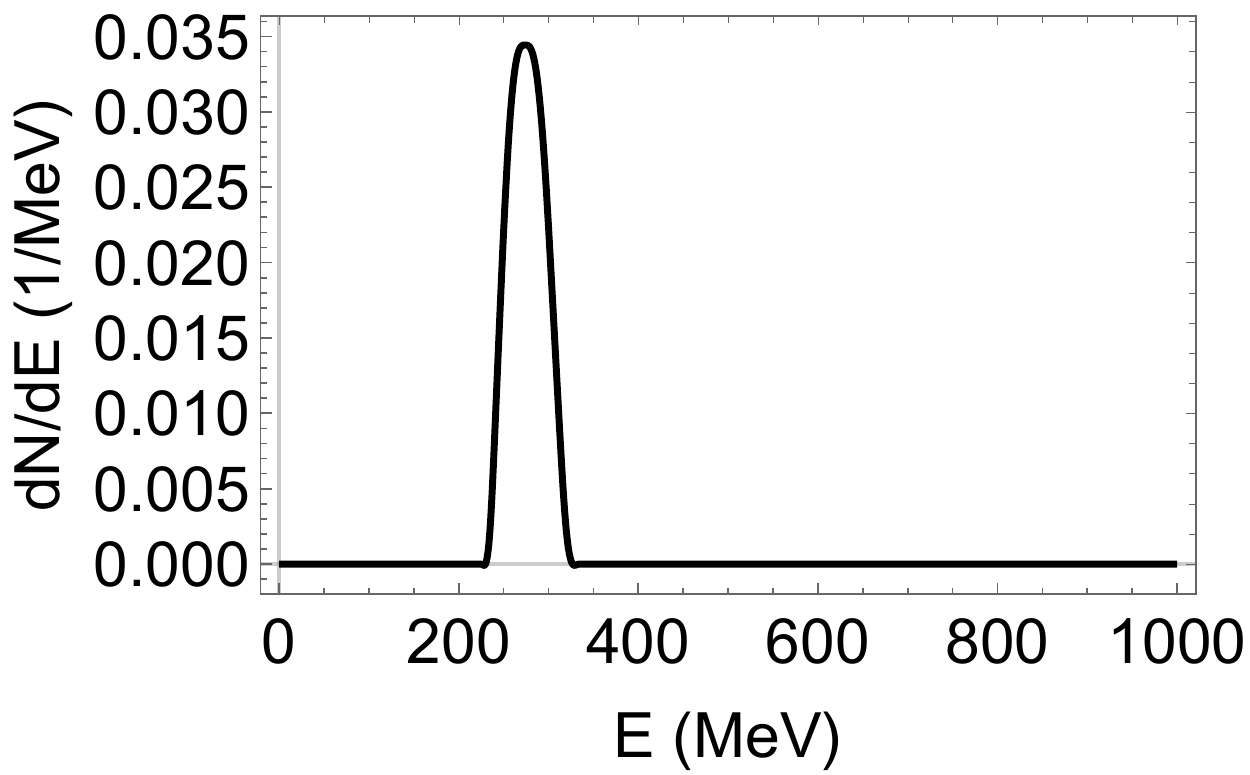}
  \caption{The photon spectrum $dN_\gamma / dE_\gamma$ arising from the decay process
  $\eta \rightarrow \gamma \gamma$, as
  a function of $E_\gamma$, assuming a final state $\eta \pi \pi$ with $\sqrt{s}=850~\mev$.  The
  integral of the spectrum is normalized to 2.  }
  \label{fig:Spectrum850}
\end{figure}

On the other hand, if a three-body final state is produced, then
the photon spectrum will be peaked at $(m_\eta/2)$  and will fall off monotonically in either direction
until it vanishes at the limits $(m_\eta/2) [\gamma_\eta (1\pm \beta_\eta)]$, where $\gamma_\eta$ is the maximum boost
factor for the $\eta$ which is kinematically allowed.  Following the results in~\cite{Boddy:2016hbp}, we see that the fact that
$dN_\eta / dx_1$ vanishes as $x_1 \rightarrow 2m_\eta / \sqrt{s}$ implies that the maximum in the photon spectrum at
$E_\gamma = m_\eta/2$ is a smooth peak.  For illustration, we plot in Figure~\ref{fig:Spectrum850}
the photon spectrum arising from
the decay process $\eta \rightarrow \gamma \gamma$ for a $\eta \pi \pi$ final state with $\sqrt{s} =850~\mev$.

\section{Sensitivity}

We will consider the sensitivity to dark matter annihilation/decay which can be provided by experiments which are
sensitive to ${\cal O}(10-1000)~\mev$ gamma rays.  Following~\cite{Boddy:2015efa}, we consider constraints
arising from diffuse emission in the Galactic halo, and from emission from one particular dwarf (Draco).
One can also consider limits from the Galactic Center, and these have been considered in a related context in~\cite{Bartels:2017dpb}.
However, there is great systematic uncertainty regarding astrophysical foregrounds/backgrounds from the direction of the
Galactic Center.  As a result, we do not consider this target in this work.

Since we are considering experiments which are still in the design phase, we will make no serious attempt to
optimize our analysis strategy.  Instead, we will simply choose some reasonable cuts which will give us a good estimate for
the sensitivity which can be obtained.  We will consider, as a benchmark, an experiment with an exposure of $3000~\cm^2~\yr$ and
a fractional $1-\sigma$ energy resolution of $\epsilon = 0.3$ throughout the entire range of interest.  We will also assume
an angular resolution of $\lesssim 1^\circ$ throughout the energy range of interest.  The smallest target we will consider here
is Draco, with an angular size of $1.3^\circ$, and we assume that the angular resolution is smaller than this size.

The differential photon flux arising from dark matter annihilation or decay can be written as
\bea
\frac{d^2 \Phi^{ann., dec.}}{d\Omega~dE_\gamma} &=& \frac{\Xi^{ann., dec.}}{4\pi m_X} {\bar J}^{ann., dec.} \frac{dN_\gamma}{dE_\gamma} ,
\eea
where
\bea
\Xi^{ann.} &=& \frac{\langle \sigma_A  v \rangle}{2m_X} ,
\nonumber\\
\Xi^{dec.} &=& \Gamma ,
\eea
and $\bar J^{ann., dec.}$ is the average $J$-factor of the target for either annihilation or decay (we assume that the dark matter
particle is its own anti-particle).  We consider diffuse emission in the
region $|b| > 20^\circ$, where $b$ is the latitude in Galactic coordinates.  In this region, we will take the averaged $J$-factors
to be given by~\cite{Cirelli:2010xx}
\bea
\bar J_{dif.}^{ann.} &=& 3.5 \times 10^{21}~\gev^2 ~\cm^{-5} \sr^{-1} ,
\nonumber\\
\bar J_{dif.}^{dec.} &=& 1.5 \times 10^{22}~\gev ~\cm^{-2} \sr^{-1} .
\label{eq:JFactorDiffuse}
\eea
For Draco, we will take the averaged $J$-factors to be~\cite{Geringer-Sameth:2014yza}
\bea
\bar J_{Draco}^{ann.} &=& 6.94 \times 10^{21} ~\gev^2 ~\cm^{-5} ~\sr^{-1} ,
\nonumber\\
\bar J_{Draco}^{dec.} &=& 5.77 \times 10^{21} ~\gev ~\cm^{-2} ~\sr^{-1} ,
\label{eq:JFactorDraco}
\eea
with uncertainties estimated at $\sim 60\%$~\cite{Geringer-Sameth:2014yza}.
The isotropic flux observed by COMPTEL ($0.8-30~\mev$) and EGRET (30~\mev-1~\gev)~\cite{Strong:2004de} can be well fit~\cite{Boddy:2015efa} to
the function
\bea
\frac{d^2 \Phi^{iso.}}{d\Omega~dE_{obs.}} &=& 2.74 \times 10^{-3} \left( \frac{E_{obs.}}{\mev} \right)^{-2.0} \cm^{-2} \s^{-1} \sr^{-1} \mev^{-1}.
\eea

We will assume that the energy spectrum actually reported by the
experiment is given by
\bea
\frac{dN_\gamma}{dE_{obs.}} &=& \int_0^\infty dE_\gamma ~ \frac{dN_\gamma}{dE_\gamma} R_\epsilon (E_{obs.},E_\gamma) ,
\eea
where
\bea
R_\epsilon (E_{obs.},E_\gamma) &=& \frac{1}{\sqrt{2\pi} \epsilon E_\gamma}
\exp \left(-\frac{(E_{obs.}-E_\gamma)^2}{2\epsilon^2 E_\gamma^2} \right) ,
\eea
is a smearing function which accounts for the fractional energy resolution $\epsilon$ of the instrument.

If we denote by $I_{exp.}$ the exposure of the instrument, and by $\Delta \Omega$ the solid angle viewed, then the number of events due to
dark matter annihilation or decay expected to be observed within the energy window $E_- \leq E_{obs.} \leq E_+$ is given by
\bea
N_{S} &=& \frac{\Xi^{ann., dec.}}{4\pi m_X} {\bar J}^{ann., dec.} (I_{exp.} \Delta \Omega)  \int_{E_-}^{E_+} dE_{obs.} \frac{dN_\gamma}{dE_{obs.}},
\eea
while the number of expected  events actually observed (based on the fit to COMPTEL and EGRET) is given by
\bea
N_{O} &=& 8.6 \times 10^{4} \left(\frac{\mev}{E_-} - \frac{\mev}{E_+} \right)  \frac{(I_{exp.} \Delta \Omega)}{\cm^2~\yr~\sr} .
\eea

\subsection{Bounds from diffuse emission}

Sensitivity to diffuse emission from dark matter annihilation or decay is largely controlled by systematic uncertainties in the
astrophysical background.  We will adopt the following criterion for estimating the sensitivity of a given instrument to diffuse
emission: a model can be excluded if, within any energy bin of size set by the energy resolution ($E_+ - E_- = \epsilon(E_+ + E_-)$),
$N_S > \alpha N_O$, where $\alpha$ is a constant set by the systematic uncertainty of the background.  For example, a conservative bound
would arise from setting $\alpha \sim 1$, and would be appropriate if one had little confidence in any background model; in this case, a
model could only be excluded if there was an energy bin in which the estimated number of dark matter events exceeds the entire number of
observed events (including statistical uncertainty).
If one were confident that the background were smooth, then $\alpha$ would instead be determined by small fluctuations
which could be accommodated by the uncertainty in the fit to the observed flux.  For relatively large spectral features, this uncertainty
would yield $\alpha \sim 0.15$~\cite{Boddy:2015efa,Bartels:2017dpb}, but would decrease to $\sim 0.02$ for narrow features~\cite{Bartels:2017dpb}.

This analysis then yields the following constraint:
\bea
\frac{\Xi^{ann., dec.}}{\s^{-1}} \frac{\bar J_{dif.}^{ann., dec.}}{\gev \cm^{-2} \sr^{-1}} \leq \alpha (3.4 \times 10^{-5} )   \left(\frac{2\epsilon}{1-\epsilon^2}\right)
\left[\frac{E_0}{m_X} \int_{E_0(1-\epsilon)}^{E_0(1+\epsilon)} dE_{obs.} \frac{dN_\gamma}{dE_{obs.}} \right]^{-1} ,
\eea
where $E_0$ is the center of the energy bin.
For a two-body final state, each meson produces a secondary photon spectrum which is constant over some energy range.  In this
case, $E_0$ should be chosen so that the upper edge of the energy bin ($E_0 (1+\epsilon)$) lies at the highest energy such that the
secondary photon spectrum is non-vanishing.  For a three-body final state, one should choose $E_0 = m_\eta/2$.

Note that the sensitivity of an experiment to diffuse emission is determined by the signal-to-background ratio, and is
thus independent of the exposure and angular resolution.  Near threshold, when the spectral features are sharp, sensitivity
will scale as $\epsilon^{-1}$.  But as $\sqrt{s}$ increases and the spectral features become large compared to the bin size,
the dependence on $\epsilon$ disappears.  In particular, at this level, future experiments would give no improvement over
current bounds from the EGRET flux measurement.

But it is important to note that the exposure and angular resolution of future experiments can indirectly affect
sensitivity to diffuse emission.
We have assumed an isotropic flux equal to that observed by COMPTEL and EGRET.  But if a future experiment is able to resolve
a significant number of point sources which can then be subtracted from the isotropic background, the remaining background may
be significantly smaller.  Since sensitivity to diffuse emission is determined by the signal-to-background ratio, any reduction
in the normalization of the observed isotropic flux (assuming the same spectral shape), will improve sensitivity by the same
factor.

Note also that we have not included any uncertainty in the $J$-factor.  But since $N_S$ scales linearly with $\bar J$, any deviation
of the actual $J$-factor from our estimate will simply rescale our bound by the same factor.

\subsection{Bounds from dSphs}

Sensitivity to emission from a dwarf spheroidal galaxy, on the other hand, is largely controlled by statistical uncertainties.
In this case, the observed isotropic flux can be treated as an estimate for the background in a search for emission from the
dwarf (a more accurate estimate can be made for any particular dSph by measuring the flux from the direction of the dSph, but
slightly off-axis~\cite{GeringerSameth:2011iw,Mazziotta:2012ux,GeringerSameth:2014qqa,Boddy:2018qur,Albert:2017vtb}).
This background includes all emission from dark matter
annihilation or decay outside the dwarf but
along the line of sight, as well as emission from astrophysical processes.  Given an estimate for the expected background, and
a measurement of the number of photons seen from the direction of a dSph, one can determine if any particular model is consistent
with the data to any particular statistical confidence level.  We will adopt the following criterion for estimating the sensitivity
of a given instrument to emission from a dSph: if the number of events observed from the dSph in some energy range is equal to the
expected number of
background events ($N_O$), then a model can be ruled at confidence level $n-\sigma$ if $N_S > n \sqrt{N_O}$, where $N_S$ is the expected
number of signal events in the same energy range.

We can express this constraint as
\bea
\frac{\Xi^{ann., dec.}}{\s^{-1}} \frac{\bar J_{Draco}^{ann., dec.}}{\gev \cm^{-2} \sr^{-1}} &\leq& n (1.2 \times 10^{-7} )
\left(\frac{m_X}{\mev} \right)
\left(\frac{2\epsilon}{1-\epsilon^2} \frac{\mev}{E_0}  \right)^{1/2}
\left[\int_{E_0(1-\epsilon)}^{E_0(1+\epsilon)} dE_{obs.} \frac{dN_\gamma}{dE_{obs.}} \right]^{-1}
\nonumber\\
&\,& \times \left(\frac{I_{exp.} \Delta \Omega}{\cm^2~\yr~\sr} \right)^{-1/2} ,
\eea
where $E_0$ is again the center of the energy bin.  Here, we see that sensitivity scales with
$(I_{exp.} \Delta \Omega)^{1/2}$.

\subsection{Results}

\begin{figure}[t]
  \centering
  \includegraphics[scale=0.45]{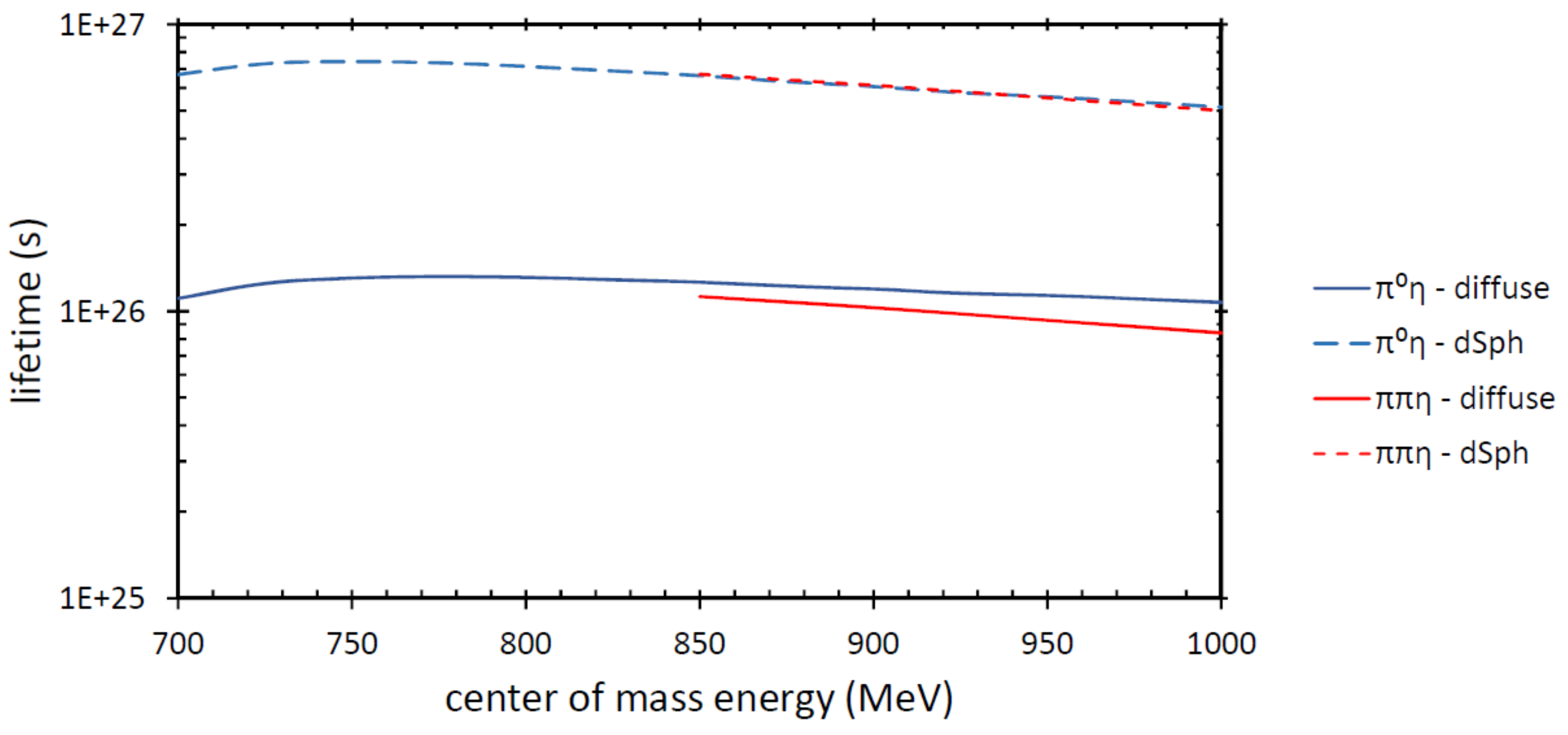}
  \caption{We plot lower bounds on the dark matter lifetime for the case in which
  the final state is $\pi^0 \eta$ (blue) or $\pi \pi \eta$ (red).  We plot conservative
  limits on diffuse emission (solid) and $2\sigma$-limits on emission from Draco (dashed).  }
  \label{fig:LifetimeBounds}
\end{figure}

\begin{figure}[t]
  \centering
  \includegraphics[scale=0.45]{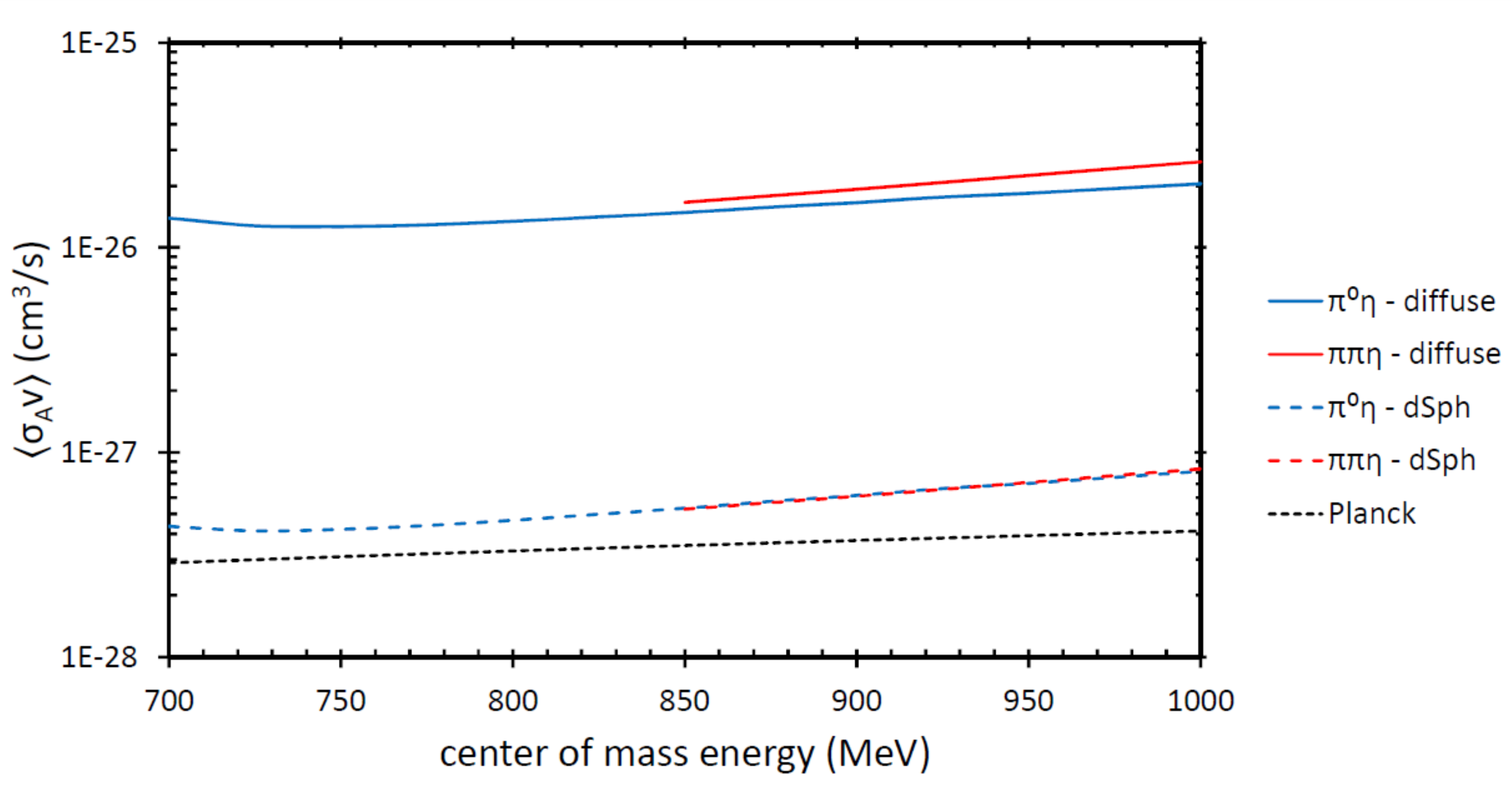}
  \caption{We plot upper bounds on the dark matter annihilation cross section for the case in which
  the final state is $\pi^0 \eta$ (blue) or $\pi \pi \eta$ (red).  We plot conservative
  limits on diffuse emission (solid) and $2\sigma$-limits on emission from Draco (dashed).
  We also plot bounds arising from Planck data~\cite{Aghanim:2018eyx} (black dotted).}
  \label{fig:CrossSectionBounds}
\end{figure}

In Figure~\ref{fig:LifetimeBounds}, we present lower bounds on the lifetime of decaying dark matter, for the case of either
a $\pi^0 \eta$ (blue) final state arising from a scalar interaction ($s^{M_2}$), or a $\pi \pi \eta$ (red) final state, arising
from a spurion which is either pseudoscalar ($p^{M_1,M_3}$), or the timelike component of an axial vector ($a_0^{M_3}$).  Solid lines denote conservative
bounds on ($\alpha=1$) diffuse photon emission, and dashed lines denote $2\sigma$-bounds on photon emission from Draco.
For the case in which dark matter
annihilates, we present similar upper bounds on the annihilation cross section in Figure~\ref{fig:CrossSectionBounds}.
In this figure, we also present recent bounds from Planck~\cite{Aghanim:2018eyx} (dotted black)  on $f_{eff.} \langle \sigma _A v \rangle /m_X$, where we
have chosen $f_{eff.} \sim 0.4$ as a rough approximation over the energy range and final states of interest~\cite{Slatyer:2015jla}.   Note,
Planck bounds on decaying dark matter only provide a lower bound on the lifetime of $\sim {\cal O}(10^{24}\s)$, which does not appear in Figure~\ref{fig:LifetimeBounds}.
For simplicity, in both figures we plot on the horizontal axis the center-of-mass energy ($\sqrt{s}$) of the process.  This is
equal to the dark matter mass in Figure~\ref{fig:LifetimeBounds}, and equal to twice the dark matter mass in Figure~\ref{fig:CrossSectionBounds}.
Since we have only included the photons arising directly from $\eta$ decay, we will have missed a small number of photons
arising from $\pi^0$ decay which are energetic enough to overlap the energy bins we consider; our bounds are thus conservative.

Although we have chosen specifications for an instrument which would roughly match the design of e-ASTROGAM, we have noted
that the sensitivity to diffuse emission is largely independent of the exposure and angular resolution.  As such, these conservative bounds
on diffuse emission represent actual bounds which one can place on dark matter models using data from EGRET.  Moreover, the sensitivity
scales directly with the diffuse $J$-factor, and inversely with the magnitude of the background.  If there are new estimates of the $J$-factor,
or if point source measurements lead to a reduction of the diffuse background flux, then these limits can be rescaled appropriately.

The sensitivity from dSphs, on the other hand, scales directly with the $J$-factor, but only with the background to the $-1/2$ power.  Moreover,
the sensitivity from dSphs scales as the square root of the exposure.  Unfortunately there is no dSph analysis available from EGRET, but for
any future experiment with any exposure, the estimated sensitivity can be derived from these results by an appropriate rescaling.
In fact, the estimated sensitivity for the benchmark case we consider is slightly worse than current bounds from Planck.  However, if the
$J$-factor for Draco is a little larger than the estimate we have used, then indirect detection could be competitive with current Planck
bounds.  Moreover, a stacked analysis of many dSphs would surely increase the sensitivity of indirect detection methods.

We note here that, in the case where dark matter annihilates through a scalar interaction to a $\pi^0 \eta$ final state,
the annihilation cross section is $p$-wave suppressed.  In this case, the astrophysical factor associated with dark matter
annihilation is not the standard $J$-factor associated with velocity-independent dark matter annihilation, but instead depends on
the dark matter velocity-distribution~\cite{Robertson:2009bh,Ferrer:2013cla,Boddy:2017vpe,Boddy:2018ike}.
A determination of the effective $J$-factor for Draco in the case of $p$-wave suppressed
dark matter annihilation is beyond the scope of this work; instead, the associated velocity-suppression is absorbed into the
quantity $\langle \sigma_A v \rangle$, which implicitly includes a convolution of the annihilation cross section with dark matter
velocity distributions in Draco.  In this case, the bound from Planck cannot be directly applied, since the typically velocity of
dark matter particles in the early Universe was far different that in the current epoch.

In the case of either dark matter decay or dark matter annihilation, the sensitivity arising from a search for emission from
Draco exceeds the sensitivity of a search for diffuse emission.   But this improvement is more pronounced for the
case of dark matter annihilation, as in this case the photon signal scales as the square of the dark matter density in the dwarf.  Similarly,
bounds from Planck are the most constraining for the case of dark matter annihilation; this improvement is driven by the fact
that the annihilation rate scales as the square of the density, which was much larger at the time of recombination than in the
present epoch.

By comparison, we can consider the prospects for a search for dark matter decay or annihilation in the Galactic Center,
which we define by the range $|b| < 5^\circ$, $|\ell| < 30^\circ$.
In this region of the sky, the observed flux is about a factor of 20 larger than
the observed diffuse flux at ${\cal O}(100-1000~\mev)$~\cite{Bartels:2017dpb}.  The average
$J$-factor for decay for the Galactic Center is a factor of 3-4 larger than for the Galactic halo at high latitudes~\cite{Cirelli:2010xx}.
As a result, the sensitivity of an instrument to dark matter decay in the GC would be worse than from diffuse emission
due to decay throughout the halo.  On the other hand, the average $J$-factor for annihilation for the GC is about a factor
of 30 larger than for the Galactic halo at high latitudes~\cite{Cirelli:2010xx}, assuming a Navarro-Frenk-White (NFW) profile.  
Thus, there is little to be gained in
searching the GC for dark matter annihilation or decay, in a conservative analysis.  Note, however, that in a related context,
the authors of~\cite{Bartels:2017dpb} found much better prospects for studying emission from sub-GeV dark matter annihilation in
the GC.  We believe that this discrepancy is largely due to the fact that we have assumed a conservative
analysis, in which a model can only be excluded if the expected number of signal events exceeds the observed number of events.
More optimistic assumptions about one's ability to understand the backgrounds from the GC were made in~\cite{Bartels:2017dpb}.

\subsection{Constraints from the LHC and Direct Detection}

If dark matter couples to light quarks, then one might hope to constrain these scenarios using
LHC mono-anything searches for processes such as
$pp \rightarrow XX~+~jet$~\cite{Aaboud:2017phn,Sirunyan:2017jix}.
But if dark matter decays to light mesons with a lifetime of $\gtrsim 10^{26}\s$, then the couplings would be so small that
LHC searches are completely unconstraining.  If
dark matter instead annihilates to light mesons, then LHC searches could become relevant, but
they are nevertheless challenging.  For sub-GeV dark matter, there is no real reason
to expect the contact approximation to be valid at ${\cal O}(\tev)$ energies; if the mass of the mediating particle
is significantly smaller than the LHC energy range, then LHC mono-anything constraints become very model-dependent,
and can easily be evaded.
The reason is that, for a fixed value of $\alpha / \Lambda^2$, a smaller mediator mass scale corresponds to a smaller
coupling.  But if the LHC energy scale is much larger than the mediator mass scale, then the LHC mono-anything production rate
becomes largely independent of the mediator mass.  But the reduced coupling then yields an overall suppression
of the LHC production cross section.
For example, if a dark matter bilinear coupling to light quarks is represented by a $p^1$ spurion, then one would
need $\alpha_P^1 / \Lambda^2 \sim ({\cal O}(100)~\gev)^{-2}$ in order for the annihilation cross section to be
${\cal O}(10^{-2})~\pb$, which is the approximate limit obtained from Planck data.  One could consider a specific example
of such a model in which dark matter coupled to quarks through the exchange of a new pseudoscalar particle with mass $\sim {\cal O}(\gev)$
and $\alpha \sim 10^{-4}$.
Such a model is not constrained by current LHC searches~\cite{Sirunyan:2017jix}, since the mono-anything production cross section
would be several orders of magnitude below current sensitivity.

Direct detection experiments, such as CRESST~\cite{Petricca:2017zdp}, are now probing the mass range we consider here.
If a dark matter bilienar interacts with quarks through a pseudoscalar interaction, then the dark matter-nucleon
scattering cross section is $v^4$-suppressed and spin-dependent, and CRESST would be unlikely to see a signal.  If a dark matter
blinear instead couples through a scalar interaction, then dark matter can also have velocity-independent
spin-independent scattering with nuclei.
But since dark matter annihilation through a scalar interaction is $p$-wave suppressed, one needs
$\alpha_S^2 / \Lambda^2 \sim ({\cal O}(1-10)~\gev)^{-2}$ in order for the annihilation cross section to be
${\cal O}(10^{-2})~\pb$.  For such models, the dark matter-proton scattering cross section would be much
larger than a picobarn.  However, for the scalar interaction generated by spurion $s^2$, dark matter interactions
are maximally isospin-violating~\cite{Chang:2010yk,Feng:2011vu,Feng:2013fyw},
which would suppress the naive sensitivity of CRESST to these models.
Such models could also be probed by the effect of scattering on CMB~\cite{Gluscevic:2017ywp,Boddy:2018kfv}, but current bounds are not constraining.  If dark matter couples to light quarks, then new nucleon-nucleon forces can be induced by one-loop diagrams
with dark matter running in the loop, and these forces can be probed by meson spectroscopy and neutron scattering experiments~\cite{Fichet:2017bng}.  But current bounds again do not constrain models of interest for us.
A detailed study of the direct detection prospects for these models for current and upcoming direct detection experiments would be
very interesting, but is beyond the scope of this work.

\section{Conclusions}

We have considered the indirect detection of sub-GeV dark matter annihilation or decay.  If dark matter couples
to quarks, then the hadronic final states and branching fractions are largely determined by symmetry and kinematics,
and can be derived in chiral perturbation theory.  In particular, striking photon signals can be produced by the
process $\eta \rightarrow \gamma \gamma$.  Especially for the case of dark matter decay, the current lower bounds
which can be obtained from EGRET data already exceed those obtained from Planck by orders of magnitude.  Future data
from an experiment such as e-ASTROGAM, looking at dwarf spheroidal galaxies, can provide an even greater improvement.

In this work, we have utilized chiral perturbation theory at lowest order, and have focused on the photons arising
directly from the decay process $\eta \rightarrow \gamma \gamma$, since the signal is well understood and the background
is small.  But dark matter annihilation or decay in this energy range generally produces a larger number of pions,
especially after including $\eta$ decay, but the kinematics are more difficult.
More generally, for slightly larger energies, a much wider range of final states is accessible and becomes relevant
for indirect detection.
But as increasing interest is shown in
sub-GeV dark matter, it would be interesting to perform a more comprehensive study of the hadronic final states which
can be produced.

In a similar vein, it is worth noting that if sub-GeV dark matter couples primarily to light quarks, then it can
potentially be produced at proton beam fixed-target or beam-dump experiments, such as NA62~\cite{NA62}
or SeaQuest~\cite{Berlin:2018pwi}, or proposed experiments such as
SHiP~\cite{Anelli:2015pba}, or related proposed experiments such as FASER~\cite{Feng:2017uoz}.
In particular, one might hope that dark matter could be produced in the rare decays of heavier mesons.  But to determine
the available signals at such experiments, a more detailed study beyond lowest order in chiral perturbation theory would be
necessary.

\section*{Acknowledgments}

We are grateful to Dillon Berger, Rebecca Leane, Arvind Rajaraman, Xerxes Tata and Mark Wise for useful discussions.
We acknowledge the organizers of the 2018 Santa Fe Summer Workshop in Particle Physics and of SUSY2018, where part of
this work was conducted, for their hospitality.
This research is funded in part by DOE grant DE-SC0010504.


\begin{thebibliography}{99}

\bibitem{Dolgov:1980uu}
  A.~D.~Dolgov,
  Yad.\ Fiz.\  {\bf 31}, 1522 (1980).

\bibitem{Carlson:1992fn}
  E.~D.~Carlson, M.~E.~Machacek and L.~J.~Hall,
  Astrophys.\ J.\  {\bf 398}, 43 (1992).
  doi:10.1086/171833

\bibitem{Moroi:1993mb}
  T.~Moroi, H.~Murayama and M.~Yamaguchi,
  Phys.\ Lett.\ B {\bf 303}, 289 (1993).
  doi:10.1016/0370-2693(93)91434-O


\bibitem{Hall:2009bx}
  L.~J.~Hall, K.~Jedamzik, J.~March-Russell and S.~M.~West,
  JHEP {\bf 1003}, 080 (2010)
  doi:10.1007/JHEP03(2010)080
  [arXiv:0911.1120 [hep-ph]].

\bibitem{Hochberg:2014dra}
  Y.~Hochberg, E.~Kuflik, T.~Volansky and J.~G.~Wacker,
  Phys.\ Rev.\ Lett.\  {\bf 113}, 171301 (2014)
  doi:10.1103/PhysRevLett.113.171301
  [arXiv:1402.5143 [hep-ph]].

\bibitem{Kuflik:2015isi}
  E.~Kuflik, M.~Perelstein, N.~R.~L.~Lorier and Y.~D.~Tsai,
  Phys.\ Rev.\ Lett.\  {\bf 116}, no. 22, 221302 (2016)
  doi:10.1103/PhysRevLett.116.221302
  [arXiv:1512.04545 [hep-ph]].

\bibitem{DeAngelis:2017gra}
  A.~De Angelis {\it et al.} [e-ASTROGAM Collaboration],
  arXiv:1711.01265 [astro-ph.HE].

\bibitem{Caputo:2017sjw}
  R.~Caputo {\it et al.} [AMEGO Team],
  PoS ICRC {\bf 2017}, 910 (2017).
  doi:10.22323/1.301.0910

\bibitem{Boddy:2015efa}
  K.~K.~Boddy and J.~Kumar,
  Phys.\ Rev.\ D {\bf 92}, no. 2, 023533 (2015)
  doi:10.1103/PhysRevD.92.023533
  [arXiv:1504.04024 [astro-ph.CO]].

\bibitem{Boddy:2015fsa}
  K.~K.~Boddy and J.~Kumar,
  AIP Conf.\ Proc.\  {\bf 1743}, 020009 (2016)
  doi:10.1063/1.4953276
  [arXiv:1509.03333 [astro-ph.CO]].

\bibitem{Boddy:2016fds}
  K.~K.~Boddy, K.~R.~Dienes, D.~Kim, J.~Kumar, J.~C.~Park and B.~Thomas,
  Phys.\ Rev.\ D {\bf 94}, no. 9, 095027 (2016)
  doi:10.1103/PhysRevD.94.095027
  [arXiv:1606.07440 [hep-ph]].

\bibitem{Bartels:2017dpb}
  R.~Bartels, D.~Gaggero and C.~Weniger,
  JCAP {\bf 1705}, no. 05, 001 (2017)
  doi:10.1088/1475-7516/2017/05/001
  [arXiv:1703.02546 [astro-ph.HE]].

\bibitem{Cata:2017jar}
  O.~Cata, A.~Ibarra and S.~Ingenhütt,
  JCAP {\bf 1711}, no. 11, 044 (2017)
  doi:10.1088/1475-7516/2017/11/044
  [arXiv:1707.08480 [hep-ph]].

\bibitem{Dutra:2018gmv}
  M.~Dutra, M.~Lindner, S.~Profumo, F.~S.~Queiroz, W.~Rodejohann and C.~Siqueira,
  JCAP {\bf 1803}, 037 (2018)
  doi:10.1088/1475-7516/2018/03/037
  [arXiv:1801.05447 [hep-ph]].

\bibitem{CHPTReviews}
  J.~Gasser and H.~Leutwyler,
  Nucl.\ Phys.\ B {\bf 250}, 465 (1985); \\
  J.~Gasser and H.~Leutwyler,
  Annals Phys.\  {\bf 158}, 142 (1984) ; \\
  U.~G.~Meissner,
  Rept.\ Prog.\ Phys.\  {\bf 56}, 903 (1993)
  [hep-ph/9302247];\\
  G.~Ecker,
  Prog.\ Part.\ Nucl.\ Phys.\  {\bf 35}, 1 (1995)
  [hep-ph/9501357];\\
  A.~Pich,
  Rept.\ Prog.\ Phys.\  {\bf 58}, 563 (1995)
  [hep-ph/9502366];\\
  G.~Colangelo and G.~Isidori,
  hep-ph/0101264;\\
  S.~Scherer,
  Adv.\ Nucl.\ Phys.\  {\bf 27}, 277 (2003)
  [hep-ph/0210398].



\bibitem{Kumar:2013iva}
  J.~Kumar and D.~Marfatia,
  Phys.\ Rev.\ D {\bf 88}, no. 1, 014035 (2013)
  doi:10.1103/PhysRevD.88.014035
  [arXiv:1305.1611 [hep-ph]].



\bibitem{Boddy:2016hbp}
  K.~K.~Boddy, K.~R.~Dienes, D.~Kim, J.~Kumar, J.~C.~Park and B.~Thomas,
  Phys.\ Rev.\ D {\bf 95}, no. 5, 055024 (2017)
  doi:10.1103/PhysRevD.95.055024
  [arXiv:1609.09104 [hep-ph]].



\bibitem{Cirelli:2010xx}
  M.~Cirelli {\it et al.},
  JCAP {\bf 1103}, 051 (2011)
  Erratum: [JCAP {\bf 1210}, E01 (2012)]
  doi:10.1088/1475-7516/2012/10/E01, 10.1088/1475-7516/2011/03/051
  [arXiv:1012.4515 [hep-ph]].

\bibitem{Geringer-Sameth:2014yza}
  A.~Geringer-Sameth, S.~M.~Koushiappas and M.~Walker,
  Astrophys.\ J.\  {\bf 801}, no. 2, 74 (2015)
  doi:10.1088/0004-637X/801/2/74
  [arXiv:1408.0002 [astro-ph.CO]].

\bibitem{Strong:2004de}
  A.~W.~Strong, I.~V.~Moskalenko and O.~Reimer,
  Astrophys.\ J.\  {\bf 613}, 962 (2004)
  doi:10.1086/423193
  [astro-ph/0406254].

\bibitem{GeringerSameth:2011iw}
  A.~Geringer-Sameth and S.~M.~Koushiappas,
  Phys.\ Rev.\ Lett.\  {\bf 107}, 241303 (2011)
  doi:10.1103/PhysRevLett.107.241303
  [arXiv:1108.2914 [astro-ph.CO]].

\bibitem{Mazziotta:2012ux}
  M.~N.~Mazziotta, F.~Loparco, F.~de Palma and N.~Giglietto,
  Astropart.\ Phys.\  {\bf 37}, 26 (2012)
  doi:10.1016/j.astropartphys.2012.07.005
  [arXiv:1203.6731 [astro-ph.IM]].

\bibitem{GeringerSameth:2014qqa}
  A.~Geringer-Sameth, S.~M.~Koushiappas and M.~G.~Walker,
  Phys.\ Rev.\ D {\bf 91}, no. 8, 083535 (2015)
  doi:10.1103/PhysRevD.91.083535
  [arXiv:1410.2242 [astro-ph.CO]].

\bibitem{Boddy:2018qur}
  K.~Boddy, J.~Kumar, D.~Marfatia and P.~Sandick,
  Phys.\ Rev.\ D {\bf 97}, no. 9, 095031 (2018)
  doi:10.1103/PhysRevD.97.095031
  [arXiv:1802.03826 [hep-ph]].

\bibitem{Albert:2017vtb}
  A.~Albert {\it et al.} [HAWC Collaboration],
  Astrophys.\ J.\  {\bf 853}, no. 2, 154 (2018)
  doi:10.3847/1538-4357/aaa6d8
  [arXiv:1706.01277 [astro-ph.HE]].

\bibitem{Aghanim:2018eyx}
  N.~Aghanim {\it et al.} [Planck Collaboration],
  arXiv:1807.06209 [astro-ph.CO].

\bibitem{Slatyer:2015jla}
  T.~R.~Slatyer,
  Phys.\ Rev.\ D {\bf 93}, no. 2, 023527 (2016)
  doi:10.1103/PhysRevD.93.023527
  [arXiv:1506.03811 [hep-ph]].

\bibitem{Robertson:2009bh}
  B.~Robertson and A.~Zentner,
  Phys.\ Rev.\ D {\bf 79}, 083525 (2009)
  doi:10.1103/PhysRevD.79.083525
  [arXiv:0902.0362 [astro-ph.CO]].

\bibitem{Ferrer:2013cla}
  F.~Ferrer and D.~R.~Hunter,
  JCAP {\bf 1309}, 005 (2013)
  doi:10.1088/1475-7516/2013/09/005
  [arXiv:1306.6586 [astro-ph.HE]].

\bibitem{Boddy:2017vpe}
  K.~K.~Boddy, J.~Kumar, L.~E.~Strigari and M.~Y.~Wang,
  Phys.\ Rev.\ D {\bf 95}, no. 12, 123008 (2017)
  doi:10.1103/PhysRevD.95.123008
  [arXiv:1702.00408 [astro-ph.CO]].

\bibitem{Boddy:2018ike}
  K.~K.~Boddy, J.~Kumar and L.~E.~Strigari,
  arXiv:1805.08379 [astro-ph.HE].

\bibitem{Aaboud:2017phn}
  M.~Aaboud {\it et al.} [ATLAS Collaboration],
  JHEP {\bf 1801}, 126 (2018)
  doi:10.1007/JHEP01(2018)126
  [arXiv:1711.03301 [hep-ex]].

\bibitem{Sirunyan:2017jix}
  A.~M.~Sirunyan {\it et al.} [CMS Collaboration],
  Phys.\ Rev.\ D {\bf 97}, no. 9, 092005 (2018)
  doi:10.1103/PhysRevD.97.092005
  [arXiv:1712.02345 [hep-ex]].

\bibitem{Petricca:2017zdp}
  F.~Petricca {\it et al.} [CRESST Collaboration],
  arXiv:1711.07692 [astro-ph.CO].

\bibitem{Chang:2010yk}
  S.~Chang, J.~Liu, A.~Pierce, N.~Weiner and I.~Yavin,
  JCAP {\bf 1008}, 018 (2010)
  doi:10.1088/1475-7516/2010/08/018
  [arXiv:1004.0697 [hep-ph]].

\bibitem{Feng:2011vu}
  J.~L.~Feng, J.~Kumar, D.~Marfatia and D.~Sanford,
  Phys.\ Lett.\ B {\bf 703}, 124 (2011)
  doi:10.1016/j.physletb.2011.07.083
  [arXiv:1102.4331 [hep-ph]].

\bibitem{Feng:2013fyw}
  J.~L.~Feng, J.~Kumar and D.~Sanford,
  Phys.\ Rev.\ D {\bf 88}, no. 1, 015021 (2013)
  doi:10.1103/PhysRevD.88.015021
  [arXiv:1306.2315 [hep-ph]].

\bibitem{Gluscevic:2017ywp}
  V.~Gluscevic and K.~K.~Boddy,
  arXiv:1712.07133 [astro-ph.CO].

\bibitem{Boddy:2018kfv}
  K.~K.~Boddy and V.~Gluscevic,
  arXiv:1801.08609 [astro-ph.CO].

\bibitem{Fichet:2017bng}
  S.~Fichet,
  Phys.\ Rev.\ Lett.\  {\bf 120}, no. 13, 131801 (2018)
  doi:10.1103/PhysRevLett.120.131801
  [arXiv:1705.10331 [hep-ph]].

\bibitem{NA62}
  B.~Döbrich [NA62 Collaboration],
  arXiv:1807.10170 [hep-ex].

\bibitem{Berlin:2018pwi}
  A.~Berlin, S.~Gori, P.~Schuster and N.~Toro,
  arXiv:1804.00661 [hep-ph].

\bibitem{Anelli:2015pba}
  M.~Anelli {\it et al.} [SHiP Collaboration],
  arXiv:1504.04956 [physics.ins-det].

\bibitem{Feng:2017uoz}
  J.~L.~Feng, I.~Galon, F.~Kling and S.~Trojanowski,
  Phys.\ Rev.\ D {\bf 97}, no. 3, 035001 (2018)
  doi:10.1103/PhysRevD.97.035001
  [arXiv:1708.09389 [hep-ph]].

\end{thebibliography}
\end{document}